# High Speed Data Retrieval from National Data Center (NDC) Reducing Time and Ignoring Spelling Error in Search Key Based on Double Metaphone Algorithm

Md. Palash Uddin[1], Ashfaque Ahmed[2], Md. Delowar Hossain[3], Masud Ibn Afjal[4] and Shah Md. Tanvir Siddiquee[5]

[1]Lecturer, Department of Computer Science and Information Technology, Hajee Mohammad Danesh Science and Technology University, Dinajpur-5200, Bangladesh.

[2]Software Developer, Matrivumisoft Ltd., Dhaka, Bangladesh.

[3]Assistant Professor, Department of Computer Science and Information Technology, Hajee Mohammad Danesh Science and Technology University, Dinajpur, Bangladesh.

[4]Lecturer, Department of Computer Engineering, Hajee Mohammad Danesh Science and Technology University, Dinajpur, Bangladesh

[5]M.Sc. Student, Department of Computer Science, South Asian University, India.

## ABSTRACT

*Fast and efficient data management is one of the demanding technologies of today's aspect. This paper proposes a system which makes the working procedures of present manual system of storing and retrieving huge citizen's information of Bangladesh automated and increases its effectiveness. The implemented search methodology is user friendly and efficient enough for high speed data retrieval ignoring spelling error in the input keywords used for searching a particular citizen. The main concern in this research is minimizing the total searching time for a given keyword. This can be done if we can pre-establish the idea of getting the data belonging to the searching keyword. The primary and secondary key-code generated by the Double Metaphone Algorithm for each word is used to establish that idea about the word. This algorithm is used for creating the map of the original database, through which the keyword is matched against the data.*

## 1. INTRODUCTION

Bangladesh is a developing country. In this country most of the government organizations are operating manually. It is well-known to us that the requirement of the digital world is to replace the paper work with digital components having a good efficiency and effectiveness because from the ancient time human beings always try to ease their tasks through applying some sort of technical attempts. If the manual systems of a country are being automated possessing high efficiency, the development of the country raises. So as technology advances we need to move on. From this thinking we have decided to digitize the NDC, a storage house of national resources, emphasizing on the citizens of Bangladesh ensuring a great efficiency and effectiveness. At present the information of the citizens of Bangladesh are not stored in a web-based central database. But partially the government organizations and offices do so in a scattered manner. For this reason the People's Republic of Bangladesh has started a system on National Population Register. The Bangladesh Bureau of Statistics (BBS) has undertaken a pilot project to study the feasibility of creating the national database- the National Population Register. As it is just the beginning of the project so what could be the actual system, what will be functionalities and what will be the efficiency and effectiveness it is unknown.

## 1.1. Present System

To know about the present system related with storing and retrieving the information of the citizens of Bangladesh we visited some Union Parishad (U.P.) and Pourasava in Dinajpur, Bangladesh. Among the available tools for gathering information about the present system, we have taken interview of different employees associated with the present system. We preferred interview because of its multidimensional purposes. The collected information leads us to design the structure of the present system of national database. We realize that the stored information of the citizens is not consistent and also the information is not centralized. The information flow of the present system is shown in Figure 1.

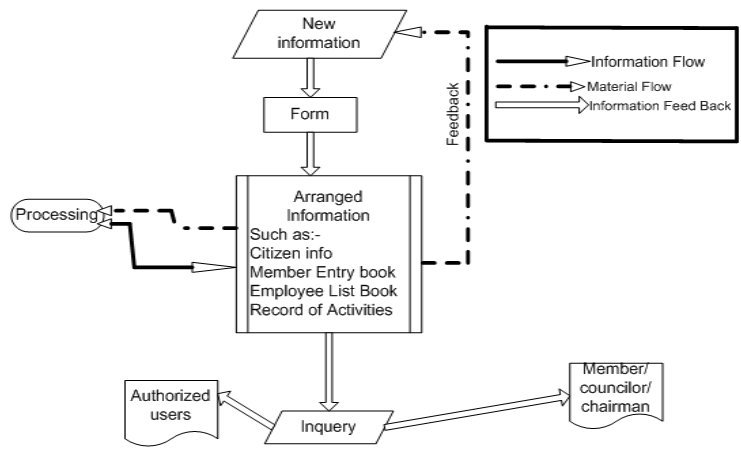

Figure 1. Present System Information Flow font.

## 1.2. Drawbacks of the Present System

The present system is a manual one where the staff members have to process everything manually. Present system faces many problems for its manual activities. Some of the problems are as follows:

- Problems in creating central database: It is very difficult to create a central database for the system as all the information is not available in database.
- Problems in retrieving data: Retrieving specific data from a catalogue is very important for the user in time of need. But the present manual system fails to show satisfactory

   performance about the information about a particular citizen. And as a result it becomes the source of hinders in searching.
- Fail to maintain unique data: The records are stored may have the possibility of duplication i.e. one person can have names or other details. Further one person can entry his/her record several times by simply filling the form.

### 1.3. Proposed System

National Data Center is the storage house of all the national resources of a country. Here we emphasis on the Human Resources, the citizens of Bangladesh. The NDC is built based as per international standards. An NDC provides on-demand data access and submission via a secure web-based interface. The main purpose of this system is to store the information of citizens of Bangladesh in a web-based central database and then to retrieve the information of the citizens from a huge database ignoring spelling error and reducing searching time using a phonetic algorithm namely Double Metaphone Algorithm. The system will help the top level management to make decisions regarding development of the country. It will provide necessary information to administrative personnel. These services have been designed to help users to get their jobs done faster and more efficiently ignoring spelling error complexity. The graphical interface brings new power and flexibility to their users. The most outstanding part of this web-based system is that the search technique is enhanced greatly. The information flow of the proposed system is shown in Figure 2.

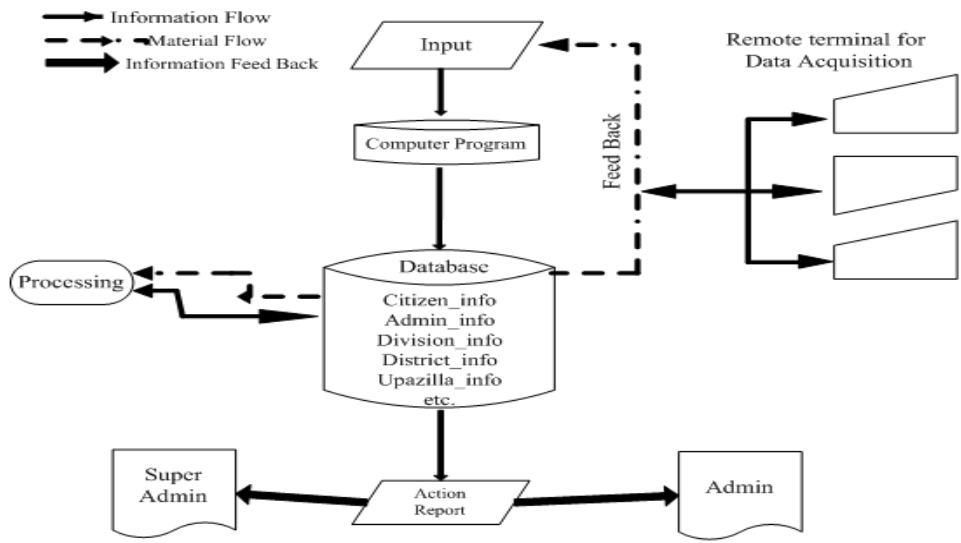

Figure 2. Proposed System Information Flow

### 1.4. Aim and Objective of the Proposed System

The automation of the system will enable the user to interact with the system in an easy way. They can search records of the citizens with ease and high speed. This will provide a fast delivery of services from the system to the user. The system will be useful also to administrative personnel to take necessary decisions regarding the development of the country. The purposes of this application are as follows:

International Journal of Computer Science, Engineering and Applications (IJCSEA) Vol.3, No.6, December 2013

- Admin can enter details related to a particular citizen.
- Super Admin can provide membership to the Admins of NDC.
- Admin can read and write information about any citizen, and can update, create, and delete the record of a particular citizen as per requirement and implementation plans.
- Admin and Super Admin can view their status and make search for citizen.
- Providing high security

## 2. METHODS AND MATERIALS

In the implementation phase of NDC we have used the Entity-Relationship model to design database that will store and organize the national data. We have stored the data in MySQL Database Server and populated it with some sample data. The system can keep track of admin info, admin accounts, records of citizen etc. Using HTML, PHP, JavaScript, and CSS we have created an Internet-based graphical user interface that allows users to access the system from any location.

### 2.1. Search Methodology

To fulfil the requirement of the implementation of NDC effectively and efficiently we have developed a search methodology which is based on Double Metaphone Algorithm [1]. The algorithm is very useful for this purpose because it eliminates the problem during search due to misspelling of given keywords and increases the speed of search by reducing the search domain based on the given keyword with the rapid increase in the amount of data. The Double Metaphone phonetic encoding algorithm is the second generation of the Metaphone algorithm. Its implementation was described in the June 2000 issue of C/C++ Users Journal by Lawrence Philips. It makes a number of fundamental design improvements over the original Metaphone algorithm [2]. It is called "Double" because it can return both a primary and a secondary code for a string; this accounts for some ambiguous cases as well as for multiple variants of surnames with common ancestry. For example, encoding the name "Smith" yields a primary code of SM0 and a secondary code of XMT, while the name "Schmidt" yields a primary code of XMT and a secondary code of SMT--both have XMT in common. It is very useful in using indexing for not only speed in searching process but also for ignoring the spelling error in the input keyword. Actually, the available choices were None, Stemmer Algorithm, Soundex [3,4] or Phonix Algorithm, Metaphone Algorithm, Double Metaphone Algorithm etc. In general, the none-technique requires maximum number of comparisons to produce a result whereas stemming minimizes comparisons compared to the first but not as big. Again Soundex algorithm generates less number of unique codes but it becomes ambiguous if the number of data is very large. Metaphone algorithm is more reliable than Soundex algorithm, but it also shows some limitations for some ambiguous words. Double Metaphone algorithm is the modifications of the Metaphone algorithm and it eliminates the limitations of Metaphone. It is comparatively more reliable than any of the above algorithm for making fuzzy indexing.

### 2.2. Related Works

Double Metaphone algorithm is mainly designed for finding the phonetic similarity between two words. It has been applied in several applications like:

- Naushad UzZaman and Mumit Khan, "A Bangla Phonetic Encoding for Better Spelling Suggestions" [5].
- Chakkrit Snae and Michael Brückner, "Novel Phonetic Name Matching Algorithm with a Statistical Ontology for Analyzing Names Given in Accordance with Thai Astrology" [6].

- A.K. Mandal, M.D. Hossain and M. Nadim, "An Efficient Search Suggestion Generator, Ignoring Spelling Error for High Speed Data Retrieval Using Double Metaphone Algorithm" [7].
- J. Zobel and P. Dart, "Finding Approximate Matches in Large Lexicons" [8].
- Naushad UzZaman and Mumit Khan, "A Double Metaphone Encoding for Approximate Name Searching and Matching in Bangla" [9].

## 2.3. Interface between Search Result and the Original Database

The interface between search result and the original database is maintained by introducing data pointers. Data pointer is a mean by which a particular row of a table can be identified. For example if we know value of the primary key of a particular table then we can easily get the value of the specific row. Thus during the searching phase it is not needed to originally search the whole database, just getting the appropriate data pointer and retrieve the original data located by the pointer. The process of the creation of the data pointer is described below:

- The entire tables whose information may be required to search are given unique id.
- Each Data Pointer comprises with table id and primary key value of the row.

The tables and id used in the database search of National Data Center are as follows:

Table 1. Defining Informative Tables in Search Methodology

| Table Name | Purpose of the Table | ID |
|---|---|---|
| citizen | Storing the information of citizen | 0 |

Thus if we get a Data Pointer (0, 133) thus following SQL query will be performed for getting the data from the table:

SELECT * FROM citizen WHERE p_value =133;

## 2.4. E-R Diagram of the Search Methodology

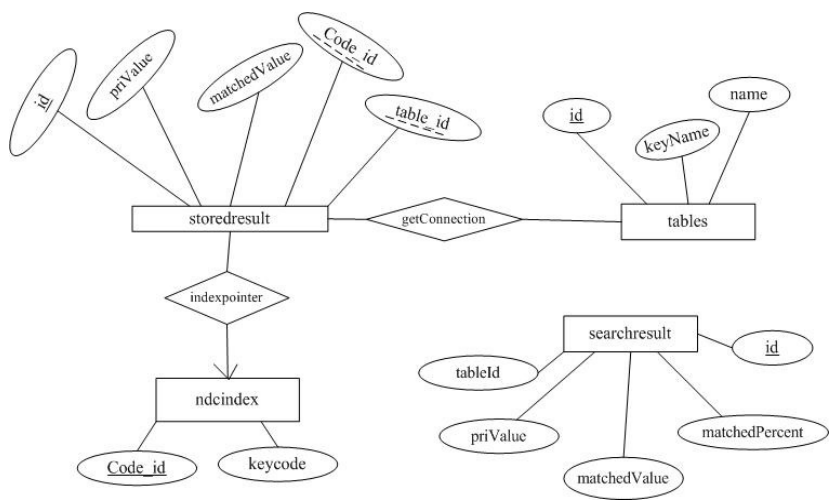

Figure 3: E-R Diagram of the Search Methodology

## 2.5. Purposes of the Data Tables Used in Search Methodology

- Table ndcindex, stores the Metaphone Code of each word, is indexed against keycode, for ensuring high speed retrieval of id.
- Table tables is defined by using two dimensional array.
- Table storedresult, stores the map of original database, is indexed against tableid and codeid for ensuring high speed navigation of exact location of the searched value.
- Table searchresult is a temporary storage of searched result, and it will be updated with every search operation.

## 2.6. State Diagram of Search Methodology

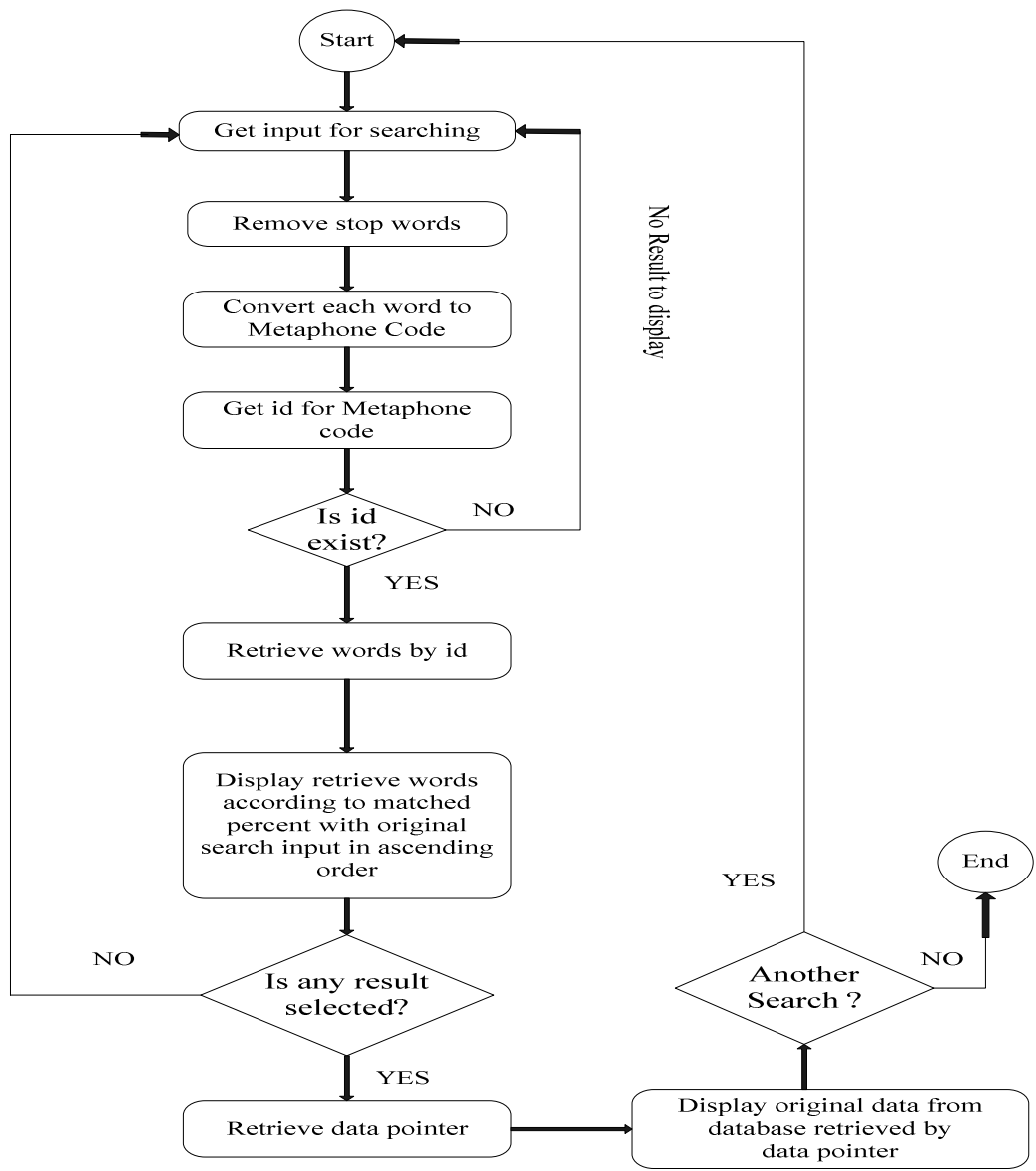

Figure 4. State Diagram of Search Methodology

## 2.6. Explanation with Example

The searching process for a particular user input (Here spelling error occurred and user desiring output for Rahim Dinajpur) will be performed by the search methodology is shown in Figure 5.

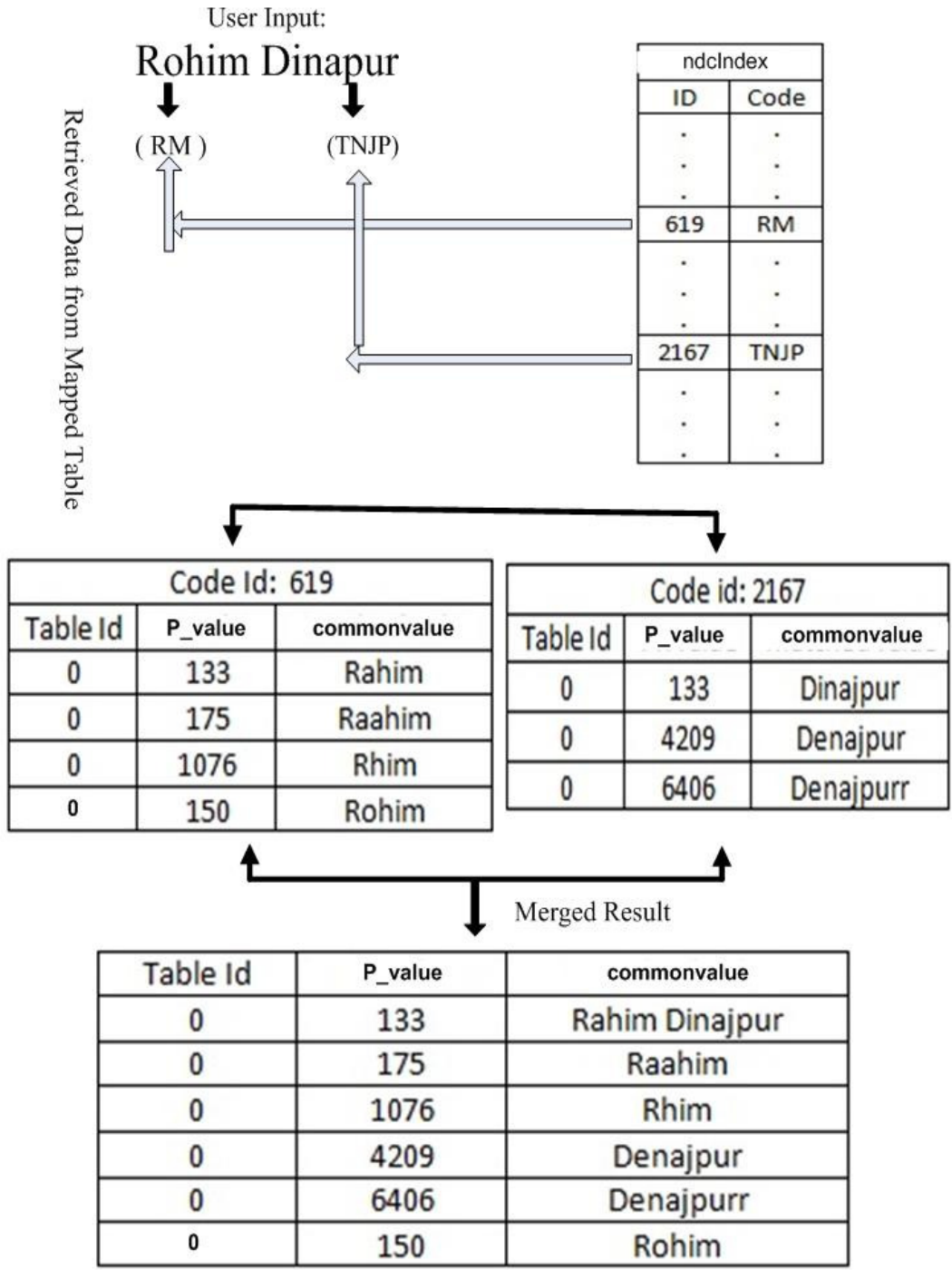

Figure 5. Explanation with Example

International Journal of Computer Science, Engineering and Applications (IJCSEA) Vol.3, No.6, December 2013

## 3. RESULT AND DISCUSSION

The simplest searching technique is the linear search which is applicable when we do not know about the data structure organization. Thus we need to search for a record sequentially and which possesses time complexity of O(n) in the worst case. However, if we know about the data structure organization that is sorted in an order, then we do not need to search sequentially. We have to apply a faster searching algorithm namely binary search technique possessing time complexity of O (logn), which is much more efficient than the linear search. But the binary search technique becomes unable to find out desired result if misspelling occurs in the input keywords. Moreover, there is no chance to apply the linear search to the same. The Double Metaphone Algorithm used to implement our search methodology shows excellent performance in such a situation when the input keywords are misspelled. Because the algorithm not only implements binary search in such situation but also reduces the search domain depending on the input keywords. We have performed an experiment by taking random citizen information, where increase of comparison is expressed with the increase of number of citizen. The amount of data versus time required to search the database is plotted here:

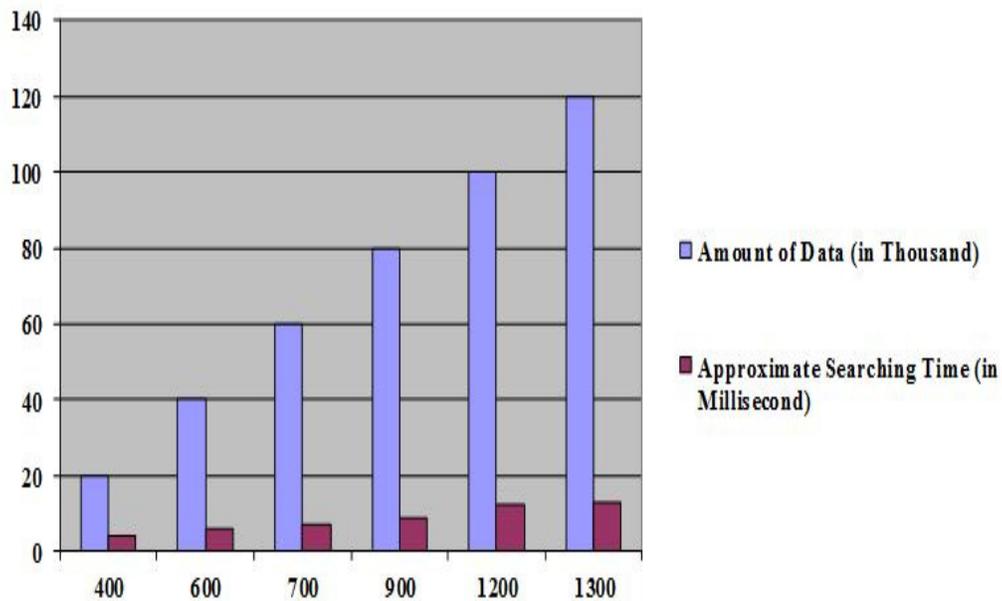

Figure 6. Amount of Data versus Searching Time Graph.

Suppose that a user wants to retrieve information about "**Abdullah Khulna**" but he/she has mistyped. Here, **Abdullah** is a person and **Khulna** is a district of Bangladesh. In this case the implemented search methodology suggests the related results as shown in Figure 6.

International Journal of Computer Science, Engineering and Applications (IJCSEA) Vol.3, No.6, December 2013| Serial No. | Matched Info | Matched (%) | More Info |
|---|---|---|---|
| Data Retrived From: ( Citizen's Record ): | | | |
| 1 | **Abdullah** , **Khulna** , Chandpur , Haimchar , Naikhong , Gorea , Employee , 8801700041114 | 97 | More |
| 2 | **Abdullah** , **Khulna** , Chandpur , Haimchar , Naikhong , Gorea , Employee , 8801700041114 | 97 | More |
| 3 | **Abdullah** , **Khulna** , Chandpur , Haimchar , Naikhong , Gorea , Employee , 8801700041114 | 97 | More |
| 4 | **Abdullah** , **Khulna** , Chandpur , Haimchar , Naikhong , Gorea , Employee , 8801700041114 | 97 | More |
| 5 | **Abdullah** , **Khulna** , Chandpur , Haimchar , Naikhong , Gorea , Employee , 8801700041114 | 97 | More |
| 6 | **Abdullah** , **Khulna** , Chandpur , Haimchar , Naikhong , Gorea , Employee , 8801700041114 | 97 | More |
| 7 | **Abdullah** , **Khulna** , Chandpur , Haimchar , Naikhong , Gorea , Employee , 8801700041114 | 97 | More |
| 8 | **Ibtihal** , Barisal , Barisal , Bakerganj , Kabai , Rabglipara , Businessman , 8801700003148 | 29 | More |
| 9 | **Ibtihal** , Barisal , Patuakhali , Kalaparaipa , Chakamaiya , Rabglipara , Teacher , 8801700012390 | 29 | More |

Figure 7. An Example of Retrieving Information Though Misspelling Occurs

Analyzing the Figure 6 and Figure 7 we can easily conclude that if the amount of data increases, the searching time increases slowly or not at all. Thus the speed of implemented search methodology which is based on Double Metaphone Algorithm is very high than the normal linear or binary search technique.

## 4. CONCLUSIONS

Since Bangladesh is a developing country its faster development totally depends on how much technologies she can use in an efficient and effective way. In other words, how fast her organizations are being digitalized with a high degree of efficiency. Bangladesh is still using the manual system for storing and retrieving the information of her citizens. There a lot of time and resources are being wasted. To save that time and resource we have tried to simulate this web-based NDC. The main goal is to replace the manual system by computerized one, which implements a faster search methodology that uses a phonetic algorithm called Double Metaphone Algorithm. The algorithm is able to trace misspelled keywords and hence the proposed system still shows relevant search results which make the system more comfortable. Using this algorithm the searching time for a particular citizen from huge stored information in the NDC has greatly reduced. In our analysis we have used advanced techniques to prepare a convenient design to make the implementation phase easy. Therefore, we hope that this system with the implemented faster and efficient search methodology will appear in a good case. Under this model framework, we can apply some artificial intelligent methods and technologies to improve the quality and effectiveness of information retrieval. In addition, this research result will be applied to the other governmental or non-governmental organizations.

## Authors

Md. Palash Uddin (palash_cse@hstu.ac.bd) received his B.Sc. degree in Computer Science and Engineering from Hajee Mohammad Danesh Science and Technology University, Dinajpur, Bangladesh in 2013. His main working interest is based on artificial intelligence, bioinformatics, algorithm analysis, database structure analysis, software engineering, theory of computation etc. Currently he is working as a lecturer in Dept. of Computer Science and Information Technology in Hajee Mohammad Danesh Science and Technology University, Dinajpur, Bangladesh. Previously, he was a lecturer in department of Computer Science and Engineering at Central Women's University, Dhaka, Bangladesh. He has research publications in various fields of Computer Science and Engineering.

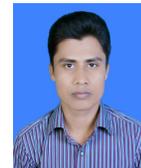

Ashfaque Ahmed is a software developer in Mutrivumisoft Ltd, Dhaka, Bangladesh. He holds a B.Sc. in Computer Science and Engineering from Hajee Mohammad Danesh Science and Technology University, Dinajpur, Bangladesh.

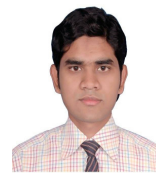

Md. Delowar Hossain is an Assistant Professor in Department of Computer science and Information Technology under the faculty of Computer Science and Engineering, Hajee Mohammad Danesh Science and Technology University, Dinajpur, Bangladesh. He holds a B.Sc. and M.Sc. in Information and Communication Engineering from Islamic University, Kushtia, Bangladesh.

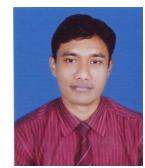

Masud Ibn Afjal is a Lecturer in Department of Computer Engineering under the faculty of Computer Science and Engineering, Hajee Mohammad Danesh Science and Technology University, Dinajpur, Bangladesh. He holds a B.Sc. in Computer Science and Engineering from Hajee Mohammad Danesh Science and Technology University, Dinajpur, Bangladesh. 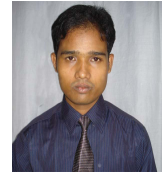

Shah Md. Tanvir Siddiquee was born in Rangpur, Bangladesh, 1988. He completed his B.Sc. degree in Computer Science and Engineering (CSE) from Hajee Mohammad Danesh Science and Technology University (HSTU), Bangladesh, in 2011. After completing Bachelor degree he worked in IT sector & worked for open source software. Now, he is pursuing M.Sc. in Computer Science at South Asian University, New Delhi, India. His research interests in Cloud Computing, Cloud Security using open source cloud environment. Previously he worked in VHDL programming, FPGA as Masters 2nd semester project. 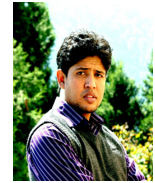